\documentclass[twocolumn,secnumarabic,amssymb, nobibnotes, aps, prd]{revtex4}
\usepackage{graphicx}

\begin{document}
\title{Why the persistent power can be observed in mesoscopic quantum system}
\author{Alexey Nikulov}
\email[]{nikulov@ipmt-hpm.ac.ru}
\affiliation{Institute of Microelectronics Technology and High Purity Materials, Russian Academy of Sciences, 142432 Chernogolovka, Moscow District, RUSSIA. }
%\date{}
\begin{abstract}It is shown that one of consequences of basic principle of
quantum mechanics at the mesoscopic level is violation of the second law of
thermodynamics and that an experimental evidence of this violation was
obtained long ago.  \end{abstract}

\maketitle

\narrowtext

It was predicted in \cite{PRB2001} that quantum oscillations of the dc
voltage in function of magnetic field $V(\Phi/\Phi_{0})$  can be observed
on segments of an inhomogeneous superconducting loop. This theoretical
prediction was published in Physical Review B  but the experimental results
\cite{Dub01} corroborating it was rejected for publication in the same
journal. Such inconsistency may be connected with a contradiction of to the
second law of thermodynamics. First of all the theoretical result
\cite{PRB2001} is challenge to the second law but about this contradiction
was openly written only in the first version \cite{0105059} of the
experimental paper rejected by Referees of Phys.Rev.B. Moreover, Referees
have rejected the Comment \cite{Comment} and Reply \cite{Reply} which point
out to the obvious contradiction of \cite{PRB2001} to the second law.

Such attitude of Referees reaffirms the words by Arthur Eddington that {\it
The second law of thermodynamics holds the supreme position among the laws
of Nature} \cite{Eddingto}. Nevertheless enough many papers with challenge
the absolute status of the second law were published in the last years
\cite{Theo,Vlada,Gordon,Daniel,Weiss}. These publications could be possible
since some experts in statistical physics are well-acquainted that the
second law does not have a fully satisfactory theoretical proof, nor has it
been fully tested experimentally. The attempts to deduce the second law
from different essential postulates are undertaken up to now
\cite{Martynov,Lieb}. Nevertheless most scientists consider for the present
any doubt in the second law as no scientific problem. Therefore I should
try once again to explain the Referees of Phys.Rev.B and other readers that
a well known phenomenon is experimental evidence of violation of the second
law and that this violation may be easy deduced from basic principle of
quantum mechanics. One may say that violation of the second law is one of
the most ordinary and obvious consequences of quantum mechanics at the
macroscopic (mesoscopic) level.

In order to understand the motive of an almost mystical view of the second
law and why its violation is obvious consequence of quantum mechanics it is
useful to recollect the history. The Carnot's principle, {\it which we call
since Clausius time the second law of thermodynamics} \cite{Smolucho}, was
based on the belief in impossibility of perpetuum mobile. This belief is
the first and sole substantiation of the second law from the Carnot's time
and up to now and the main motive of its {\it supreme position among the
laws of Nature}. Sadi Carnot wrote, in his work of a genius \cite{Carnot},
that any useful work can not be obtained from heat without a temperature
difference since such possibility means a possibility of perpetuum mobile.
It was obvious already in that time that in order the perpetuum mobile
could be impossible an irreversible behavior should be observed in nature.
Carnot wrote in \cite{Carnot} "Everyone knows that heat can produce
motion". And "everyone" knew already in that time that a cart bring to a
stop without any motive power since their kinetic energy is transformed to
the heat. It is followed from this knowledge both in that time and our time
that if this energy transformation could be irreversible then the perpetuum
mobile is inevitable. Because of this the irreversibility has become an
integral part of thermodynamics.

Both Clausius's (proposed in 1850) and Thomson's (proposed in 1851)
formulations of the second law state in fact that some thermodynamic
processes should be irreversible. The demand of the irreversibility caused
the well known collision between dynamics and thermodynamics. It is
interesting that the atomic-kinetic theory of the heat proposed by Maxwell
and Boltzmann was rejected by most scientists in the 19 century
\cite{Smolucho} because of this collision whereas in the 20 century most
scientists believed that this theory has eliminated this collision. The
probabilistic substantiation of the second law, put by J.C. Maxwell
\cite{Maxwell} into the words: {\it "the second law is drawn from our
experience of bodies consisting of an immense number of molecules"},
predominate up to now \cite{Lebowitz}. This substantiation looks very
convincing \cite{Lebowitz}. But the experience of {\it immense number of
molecules} is not enough for the defence of the absolute status of the
second law from the experience of the perpetual motion, i.e. the motion of
atoms, molecules and small Brownian particles under equilibrium condition.
It is necessary to postulate absolute randomness of any perpetual motion.
The decisive importance of this postulate is obvious, for example, from the
consideration of the ratchet/pawl combination considered by Feynman
\cite{Feynman} (and earlier by Smoluchowski \cite{Smolucho}): if the
average velocity of molecules in \cite{Feynman} could be not zero under
equilibrium condition then the second law is broken even without ratchet
and pawl.

The postulate of absolute randomness of any motion under equilibrium
condition was used both in the Maxwell - Boltzmann theory and in the
Brownian motion theory by Einstein, Smoluchowski and others. It is
important that the Brownian motion is most visual evidence not only of the
perpetual motion $v \neq 0$ but also the perpetual driving force $F_{Lan}
\neq 0$ because of nonzero friction $\gamma \neq 0$. The Langevin equation
$$m\frac{dv}{dt} + \gamma v = F_{Lan} \eqno{(1)}$$ can be used for the
description both of the Brownian motion and the motion of an automobile
\cite{NikQLSL}. The only qualitative difference results from the postulate
of absolute randomness of the Brownian motion: the average force driving an
automobile should be non-zero, whereas it was postulated that the Langevin
force $F_{Lan}$ is absolutely random and its average value equals zero
$\overline{F_{Lan}} = 0$ in all cases. But if $\overline{F_{Lan}} \neq 0$
then the Brownian motion does not differ qualitatively from the motion of
an automobile and it can be used in order to drive the automobile in
violation of the second law.

We can conclude that the postulate of absolute randomness of any
equilibrium motion saved the second law in the beginning of the 20th
century when the Brownian motion was realized as perpetual motion. This
postulate can be substantiated in the limits of classical mechanics: the
average velocity of any equilibrium motion should equal zero $\overline{v}
= 0$  since if spectrum of permitted states is continuous then for any
state with a velocity $v$ a permitted state with opposite velocity $-v$ and
the same probability $P(v^{2})$ exists, therefore $\overline{v} = \Sigma
_{per.st.} v P(v^{2}) + (-v) P(v^{2}) \equiv 0$. But according to the
quantum mechanics no all states are permitted and the average velocity of
some quantum motion can be non-zero $\overline{v} \ne 0$. Thus, according
to the well known principle of the quantum mechanics the postulate of
absolute randomness of any equilibrium motion can be incorrect.

Moreover, an enough known quantum phenomenon - the persistent current
observed at nonzero resistance $R > 0$ is experimental evidence of the
non-chaotic Brownian motion with $\overline{v} \ne 0$. The persistent
current is observed just in systems with discrete spectrum of permitted
states, because of the quantization of the momentum circulation $$\oint_{l}
dlp  = \oint_{l} dl(mv + \frac{q}{c}A) = m\oint_{l} dlv +\frac{q}{c}\Phi =
n2\pi \hbar  \eqno{(2)}$$ When the magnetic flux $\Phi $ contained within a
loop is not divisible by the flux quantum $\Phi_{0} = 2 \pi \hbar c/q$
(i.e. $\Phi \ne n\Phi_{0}$) and $\Phi  \ne  (n+0.5)\Phi _{0}$ the average
velocity $\overline{v} \ne 0$ since the spectrum of permitted states of
velocity circulation $$\oint_{l} {dlv}  = \frac{2\pi \hbar}{m}(n -
\frac{\Phi}{\Phi _{0}}) \eqno{(3)}$$ is discrete. Therefore the persistent
current $j_{p} = qn_{q}\overline{v}$, i.e. the direct current under
equilibrium conditions, was observed at numerous experiments in
superconductor \cite{little62,tink75,repeat} and even in normal metal
\cite{normal} and semiconductor \cite{semicond} nanostructures.

The observations \cite{little62,tink75,repeat,normal,semicond} of $I_{p} =
sj_{p} \neq 0$ at $R > 0$ is obvious threat to the second law since the
persistent current $I_{p}$ at non-zero power dissipation $RI_{p}^{2}$
should be maintained by a dc power source $RI_{p}^{2}$ existing under
equilibrium conditions. The possibility of this quantum phenomenon was
realized enough long ago \cite{Kulik} and enough many experts in the
mesoscopic physics are well-informed now about $I_{p} \neq 0$ observed at
$R > 0$. But they state that this equilibrium phenomenon does not threaten
the second law since no work can be extracted from the equilibrium state
(see Discussion in \cite{Aristov}). But the latter statement is one of the
consequence of the second law. Some philosophers  \cite{Callende} noted
that such circular arguments are widespread for the problem of violation of
the second law.

The dc power $RI_{p}^{2}$ can be easily used for an useful work in
contrast, for example, to the chaotic Nyquist's noise \cite{Nyquist}.
Therefore the observation the dc power under equilibrium conditions, which
can be called persistent power, is evidence of an useful perpetuum mobile
in contrast to the classical Brownian motion which may be considered as
evidence of useless perpetuum mobile, see (1). In order to save the
absolute status of the second law some scientists state that the persistent
current is no quite current. Others state that existence of a finite Ohmic
resistance for a phase coherence sample is not paradoxical when one
properly takes into account the influence of the measuring leads
\cite{Landauer}. But these statements do not correspond to the experimental
results \cite{repeat,semicond,Dub02,Dub03}. The experiment
\cite{Dub01,Dub02,Dub03} shows that the behaviour of the persistent current
in a superconducting loop with nonzero resistance does not differ
qualitatively from the one of the conventional current $I = \oint_{l}E
dl/R_{l}$ induced by the Faraday's voltage $\oint_{l}E dl =
-(1/c)d\Phi/dt$.

It is well known that a potential difference $$V =
(\langle\frac{\rho}{s}\rangle_{l_{s}} - \langle\frac{\rho
}{s}\rangle_{l})l_{s}I = (\frac{R_{ls}}{l_{s}} - \frac{R_{l}}{l})l_{s}I
\eqno{(4)}$$ should be observed on a segment $l_{s}$ of an loop with
inhomogeneous resistivity $\rho$ or section $s$ if the average resistance
along the segment $R_{ls}/l_{s} = <\rho/s>_{l_{s}} = \int_{l_{s}} dl
\rho/sl_{s}$ differs from the one along the loop $R_{l}/l = <\rho/s>_{l} =
\oint_{l} dl \rho/sl$. Just this is observed on segment of asymmetric
superconducting loop with $I_{p} \neq 0$ when its average resistance is not
equal zero and differs from the one along the loop
\cite{Dub01,Dub02,Dub03}. The persistent current $I_{p}(\Phi /\Phi_{0})
\propto  (\overline{n} - \Phi /\Phi_{0})$ is a periodical function of the
magnetic flux \cite{tink75} since the thermodynamic average value
$\overline{n}$ of the quantum number $n$ (see the relations (2,3)) is close
to an integer number $n$ corresponding to minimum energy, i.e. to minimum
$(n - \Phi /\Phi_{0})^{2}$. The quantum oscillations of the dc potential
difference $V(\Phi /\Phi_{0}) \propto I_{p}(\Phi /\Phi_{0})$ were observed
on segments of Al asymmetric loops in a narrow temperature region close to
the critical temperature, $T \approx 0.95 \div 0.995 T_{c}$
\cite{Dub01,Dub02,Dub03}. This phenomenon is described in \cite{Berger}
with help of the Josephson-junctions model. It is interesting that the like 
quantum oscillations were observed on a double point Josephson contact 
as long ago, as 1967 \cite{1967}. 

It is obvious from (4) that the dc potential difference $V(\Phi /\Phi_{0})
\propto I_{p}(\Phi /\Phi_{0})$ can be observed when the average resistance
if only of loop segment is not zero, $R_{l} > 0$. Below the fluctuation
critical region, at $T < T_{c}$, $I_{p} \neq 0$ but the resistance of loop
segments even with small section area, used in \cite{Dub01,Dub02,Dub03}, is
close to zero, see Fig.5 in \cite{Dub01}. In this temperature region the
quantum oscillations $V(\Phi /\Phi_{0})$ are observed when if only loop
segment is transferred in the resistive state by an external noise
\cite{Dub01,Dub02} or controlled external ac current $I_{ac} = \Delta I
\sin(2\pi ft)$ \cite{Dub03}. According to (4) the potential difference $V$
can be observed when both $I \neq 0$ and $R_{l} > 0$, what is not possible
in any static case: $I_{p} \neq 0$ but $R_{l} = 0$ in the closed
superconducting state and $R_{l} > 0$ but $I_{p} = 0$ in the unclosed one.
Therefore it is obvious that the quantum oscillations $V(\Phi /\Phi_{0})
\propto I_{p}(\Phi /\Phi_{0})$ can be observed only if the loop is switched
between superconducting states with different connectivity. This dynamic
state, in which the average of both values are simultaneously nonzero,
$\overline{I_{p}} \neq 0$, $\overline{R_{l}} > 0$, was called in
\cite{Dub03} dynamic resistive state.

The experimental results \cite{Dub03} show that in this state the dc
potential difference is observed both on the switched loop segment with
$\overline{R_{ln}} > 0$ and on the superconducting one with
$\overline{R_{lw}} = 0$. Moreover according to the data presented on Fig.5
of \cite{Dub03} the dc voltage can be observed because of the aspiration of
the switched segment for to return to superconducting state at $j < j_{c}$.
This paradoxical behaviour is possible since the acceleration of
superconducting pairs (with the charge $q = 2e$) in the electric field
$dp/dt =2eE = 2eV_{dc}/l_{s}$ is compensated by the momentum change because
of the quantization (2) at the closing of superconducting state when the
momentum circulation of pair changes from $2e\Phi/c$ at $v = 0$ to $n2\pi
\hbar $ \cite{PRB2001}: $n2\pi \hbar - 2e\Phi/c = 2\pi \hbar (n -
\Phi/\Phi_{0})$. The momentum change of pair because of the quantization in
a time unit at reiterated switching of the loop between superconducting
states with different connectivity was called in \cite{PRB2001} quantum
force $F_{q}$. At low switching frequency $\omega \tau _{R>0} \ll 1$,
$\oint_{l} dlF_{q}  = 2\pi \hbar ( \overline{n} - \Phi/\Phi _{0})\omega $,
where $\omega  = N_{sw}/\Theta $ is the average switching frequency,
$N_{sw}$ is a number of switching during a time $\Theta $; $\tau _{R>0} =
L/R_{l}$ is the relaxation time of the persistent current at $R_{l} > 0$.
The quantum force can not be localized in any loop segment and should be
uniform along the loop $F_{p} = \oint_{l} dlF_{q}/l $ \cite{PRB2001}. The
balance of the average forces $2eV_{dc}/l_{s} + F_{p} = 0$ acting on pair
in superconducting segment gives at $\omega \tau _{R>0} \ll 1$ the relation
between the voltage $V_{dc}$ and the frequency $\omega$ $$V_{dc} =
\frac{\pi \hbar \omega}{e}( \overline{n} - \frac{\Phi}{\Phi
_{0}})\frac{l_{s}}{l} \eqno{(5)} $$ like to the Josephson one, $V = \pi
\hbar \omega/e$ \cite{Barone}.

The dependence (5) $V_{dc} \propto  \overline{n} - \Phi/\Phi _{0}$
explains the quantum oscillations $V(\Phi/\Phi_{0})$ observed in
\cite{Dub01,Dub02,Dub03}. But it is important that they can be induced by
an external noise or an external ac current with any frequency $f$, which
can be much lower than the switching frequency $\omega $ \cite{Dub03}.
Because of superposition of the external current $I_{ac}$ and the intrinsic
persistent current $I_{p}$ the superconducting state with both connectivity
can not be stable at $(s_{w} + s_{n}) (j_{c} - I_{p}/s_{n})  < I_{ac} <
(s_{w} + s_{n}) j_{c}$ and the loop should switch between they with an
intrinsic frequency $\omega \approx 1/\tau _{eq}$ determined by the time
$\tau _{eq}$ of the relaxation to the equilibrium superconducting state
\cite{Dub03}. Because of this dynamic resistive state the asymmetry of the
current-voltage curves is observed on asymmetric loops with $s_{w} \neq
s_{n}$ at $I_{p} \neq 0$ \cite{Dub03}. Its value and sign are periodical
functions of the magnetic flux  $\Phi$ (as well as the one of $I_{p}$) and
the $V(\Phi/\Phi_{0})$ induced by $I_{ac}$ with $f \ll \omega $ look as a
consequence of rectification \cite{Dub03}.

The observation of the quantum oscillations \cite{Dub01,Dub02} is
experimental evidence of an dc power source since $W = \overline{VI_{p}}
\neq 0$ at $\Phi \neq n\Phi_{0}$ and $\Phi \neq (n+0.5)\Phi_{0}$. But these
results are not direct evidence of violation of the second law since the
loop switching can be induced by both equilibrium noise (thermal
fluctuations) and an external non-equilibrium noise and the temperature
dependence of the $V(\Phi/\Phi_{0})$ amplitude, see Fig.5 in \cite{Dub01},
shows that the quantum oscillations observed in \cite{Dub01,Dub02} are
induced rather by an external electric noise. Because, according to
\cite{Dub03}, electric noise with any frequency, from zero to a quantum
limit, can induce the $V(\Phi/\Phi_{0})$, the noise temperature
$T_{noise}(\omega ) = W_{noise}/k_{B}\Delta \omega $ should be equal the
equilibrium one $T_{noise} = T$ in whole frequency $\omega $ spectrum in
order to obtain  the direct experimental evidence of violation of the
second law from the observation of the quantum oscillations
$V(\Phi/\Phi_{0})$.

But there is no necessity at all to solve this very difficult experimental
task in order to prove experimentally violation of the second law since
this violation is proved by the numerous observations of the persistent
current $I_{p}$ at $R > 0$, which are evidence of the persistent power
$RI_{p}^{2}$. First and most reliable experimental evidence of $I_{p} \neq
0$ at $R > 0$ is the Little-Parks experiment made first in 1962 year
\cite{little62}. The explanation of this phenomenon in \cite{PRB2001} as an
ordered Brownian motion is natural since it is observed only in the
fluctuation region near $T_{c}$: above this region $R > 0$ but $I_{p} = 0$
and below $I_{p} \neq 0$ but $R > 0$.

Thus, violation of the second law not only results from basic principle of
quantum mechanics and fluctuation theory but also its experimental evidence
is known for a long time. Moreover, if we will turn from thermal
fluctuations to the quantum one we can see that this violation is one of
the most easy comprehended consequences of paradoxical principles of
quantum mechanics. The average velocity $\overline{v}$ of pairs (see (3)),
the persistent current $I_{p} = s2e\overline{n_{2e}v}$ and the dc voltage
$V_{dc}$ (5) equal zero at $\Phi = (n+0.5)\Phi_{0}$, but $\overline{v^{2}}
> 0$ \cite{tink75}, when the loop is switched between permitted states $n -
\Phi/\Phi_{0} = 1/2$ and $n - \Phi/\Phi_{0} = -1/2$ having the same
probability $P(v^{2})$ but opposite velocity. These switching in time,
taking place in non-equilibrium state \cite{Dub01,Dub02,Dub03} or because
of thermal fluctuations \cite{PRB2001,little62,tink75,repeat}, should be
accompanied at $d\Phi/dt = LdI_{p}/dt \neq 0$ by an exchange of energy with
environment. But such exchange can not be because of quantum fluctuation
notwithstanding that $I_{p} = 0$ at $\Phi = (n+0.5)\Phi_{0}$ predicted
\cite{Larkin} in this case also. It is assumed that quantum superposition
of states should take place here. But could the quantum superposition be
observed at the macroscopic level where impossibility of noninvasive
measurability is not obvious? This fundamental question is raised and
discussed in \cite{Leggett}.

\end{document}